\documentclass[conference,dvipsnames]{IEEEtran}
\IEEEoverridecommandlockouts
\usepackage{cite}
\usepackage{amsmath,amssymb,amsfonts}
\usepackage{algorithmic}
\usepackage{graphicx}
\usepackage{eurosym}
\usepackage{textcomp}
\usepackage{xcolor}
\usepackage{xargs}
\usepackage{color,listings}
\usepackage{tabularx}
\usepackage{tabularray}
\usepackage{listings}
\usepackage{caption}
\usepackage{multirow}
\usepackage{ragged2e}
\usepackage[colorinlistoftodos,prependcaption,textsize=scriptsize]{todonotes}
\def\BibTeX{{\rm B\kern-.05em{\sc i\kern-.025em b}\kern-.08em
    T\kern-.1667em\lower.7ex\hbox{E}\kern-.125emX}}

\newcommandx{\mitra}[2][1=]{\todo[linecolor=OliveGreen,backgroundcolor=OliveGreen!25,bordercolor=OliveGreen,#1]{\textbf{Mitra comments: }#2}}
\newcommandx{\pragyan}[2][1=]{\todo[linecolor=Red,backgroundcolor=Yellow!25,bordercolor=Red,#1]{\textbf{Pragyan comments: }#2}}

\newcommandx{\citeme}[2][1=]{\todo[linecolor=red,backgroundcolor=red!25,bordercolor=red,#1]{\textbf{CITE} #2}}
\newcommandx{\fillin}[2][1=]{\todo[linecolor=red,backgroundcolor=red!25,bordercolor=red,#1]{\textbf{FILL IN} #2}}

\begin{document}

%\title{Towards Leveraging Community Input: A Study of Feature Request Discussion in Open-Source Projects}
\title{Towards Better Requirements from the Crowd: Developer Engagement with Feature Requests in Open Source Software}

\author{\IEEEauthorblockN{Pragyan K C\textsuperscript{1}, Rambod Ghandiparsi\textsuperscript{1}, Thomas Herron\textsuperscript{1}, John Heaps\textsuperscript{1}, Mitra Bokaei Hosseini\textsuperscript{1}}
\IEEEauthorblockA{\textsuperscript{1}University of Texas at San Antonio, San Antonio, TX, USA\\
\textit{
[pragyan.kc, rambod.ghandiparsi, thomas.herron, john.heaps, mitra.bokaeihosseini]@utsa.edu, }
}

}

\maketitle
\begin{abstract}
%As open-source software continue to grow in popularity, effective communication around feature requests is critical for responsive and inclusive development. GitHub Issues provide a key platform where users submit requests in form of issues and developers engage with them. In this study, we conduct a grounded analysis of 393 comments from 50 randomly selected feature requests from the GitHub repositories of Mastodon and Signal, focusing specifically on how developers seek clarification. We analyze comment threads to examine whether and how developers ask clarification questions in response to user-submitted feature requests. Our goal is to understand the nature of interaction between requesters and developers: How do developers seek to clarify the feature requests? How does the interaction between developers and requests follow? What leads to developers closing a feature request thread?  We also investigate whether there exist communication practices that facilitate clearer understanding and smoother decision-making. Our findings suggest that explicit clarification is uncommon; developers usually focus on aligning with project goals rather than resolving unclear text. When clarification occurs, it emphasizes understanding user intent (maybe goal instead of intent) and feasibility, rather than technical details. By characterizing the dynamics of clarification in open-source issue trackers, this work identifies patterns that can improve user-developer collaboration and inform best practices for handling feature requests effectively.
% 
As user demands evolve, effectively incorporating feature requests is crucial for maintaining software relevance and user satisfaction. Feature requests, typically expressed in natural language, often suffer from ambiguity or incomplete information due to communication gaps or the requester’s limited technical expertise. These issues can lead to misinterpretation, faulty implementation, and reduced software quality. While seeking clarification from requesters is a common strategy to mitigate these risks, little is known about how developers engage in this clarification process in practice—how they formulate clarifying questions, seek technical or contextual details, align on goals and use cases, or decide to close requests without attempting clarification. This study investigates how feature requests are prone to NL defects (i.e., ambiguous or incomplete) and the conversational dynamics of clarification in open-source software (OSS) development, aiming to understand how developers handle ambiguous or incomplete feature requests. Our findings suggest that feature requests published on the OSS platforms do possess ambiguity and incompleteness, and in some cases, both. We also find that explicit clarification for the resolution of these defects is uncommon; developers usually focus on aligning with project goals rather than resolving unclear text. When clarification occurs, it emphasizes understanding user intent/goal and feasibility, rather than technical details. By characterizing the dynamics of clarification in open-source issue trackers, this work identifies patterns that can improve user-developer collaboration and inform best practices for handling feature requests effectively.
\end{abstract}
\vspace{-1em}

\section{Introduction}
As user demands grow and shift over time, the ability to rapidly implement enhancements and additional functionalities has become vital for software to retain its competitive edge and meet user expectations. Feature requests play a crucial role in shaping software development by enabling users to suggest new functions or enhancements based on their experiences~\cite{carreno2013analysis,oriol2018fame,dalpiaz2019re,di2016would}. 
Open-source software (OSS) platforms such as GitHub provide ``Issue Tracker" where users can submit feature requests in regards to the software and communicate with the contributors. A typical feature request begins with a user-submitted issue containing a title and a short description of the feature request that the requester intends to see in the software~\cite{kallis2021predicting}. Although feature requests provide important insights into what users need, they are expressed in natural language (NL), which makes them prone to NL defects such as ambiguity and lack of completeness. These defects occur because of communication gaps, incomplete details, or the requester’s limited technical knowledge~\cite{handbook2003contract}. Consequently, developers need to interpret and clarify feature requests, which can result in wrong assumptions, faulty implementations, and decreased software quality~\cite{zowghi2003interplay}. Without sufficient detail, developers risk misinterpreting the user's needs, potentially leading to solutions that fail to address the core problem or require costly revisions later in the development process~\cite{fitzgerald2011early}. One way to mitigate these issues is by interacting with the requester to seek clarification and ensure a shared understanding of the requested feature. Previous research has extensively explored GitHub Issue Tracker for various purposes, such as issue classification, project coordination, and community engagement~\cite{kallis2019ticket, yang2023users, izquierdo2015gila, fiechter2021visualizing, hata2022github, bissyande2013got, li2022follow}. While these studies have improved our understanding of issue types, issue management, and community interaction, the usage of conversational information within issue comments to resolve NL defects in feature requests remains largely underexplored. Understanding how clarifications are sought and provided in GitHub Issues can reveal strategies to improve the quality and implementability of feature requests in open-source projects, ultimately supporting more efficient collaboration and the development of software that better meets user needs.In our study, we aim to address the challenges associated with unclear or incomplete feature requests by investigating how developers handle such feature requests in practice. Specifically, we analyze the conversational dynamics involved in clarifying feature requests within OSS issue trackers. We seek to understand when and how developers attempt clarification, what aspects they prioritize during these interactions, and in what scenarios they choose to implement requests without clarification or close request without clarification. By characterizing these practices, our work contributes to the broader understanding of requirements evolution and user-developer collaboration in OSS projects. Our findings can inform the design of tools and guidelines that facilitate effective clarification, ultimately leading to higher-quality feature specifications, reduced implementation risks, and improved satisfaction for both developers and end users. 
The remainder of this paper is organized as follows: Section II presents background and related work; Sections III entail the approach. Sections IV and V contain results \& discussion, and threats to validity. Section VI presents the conclusion and future works.

\section{Background and Related Works}
This section provides an overview of the concepts and prior work relevant to our study. We focus on three key areas: (i) Open Source Software, (ii) Natural Language (NL) Defects in Requirements, and (iii) User-Developer Collaboration on GitHub.

\subsection{Open Source Software}
Open-source software (OSS) is characterized by publicly accessible source code, which enables developers and users alike to view, modify, and contribute to projects\cite{crowston2008free}. These projects are often sustained through voluntary contributions and community-driven development models. Unlike proprietary software development, OSS development works on collaboration and transparency, where the most effective solutions are typically those that gain community consensus and support.~\cite{crowston2008free,fellerframework,shah2006motivation,coelho2017modern}

The participatory nature of OSS promotes inclusiveness by allowing diverse voices to influence the evolution of the software. Contributors from varying geographical, cultural, and professional backgrounds can propose features, fix bugs, and contribute to documentation\cite{yamauchi2000collaboration}. This collaborative model can lead to higher software quality, as it enables peer review and rapid iteration. Moreover, the open development process can improve usability, as real users participate directly in identifying and resolving usability concerns. Numerous studies have examined these benefits, emphasizing the relationship between community effort and project success in OSS ecosystems\cite{constantino2023perceptions, yamauchi2000collaboration, ye2003toward, noll2011global}.
\subsection{Natural Language (NL) Defects in Requirements}
NL enables communication between the requester, software analyst, developer, other users, and stakeholders who may have different backgrounds, often
having little or no additional training in software development~\cite{pohl1996requirements}. However, NL requests are nonetheless prone to defects that require clarification, including ambiguity and incompleteness~\cite{handbook2003contract,ezzini2021using,pohl1996requirements}.
Ambiguity occurs when language used in requirements can be interpreted in more than one way~\cite{berry2004ambiguity}. Incomplete feature requests are those that lack crucial information needed to fully describe the intended system behavior~\cite{zowghi2003interplay}. According to Boehm~\cite{boehm1984verifying}, a complete specification must detail all aspects required for correct system functionality. Both defects create challenges, hindering validation and verification processes, and ultimately leading to bad communication between developers and clients~\cite{zowghi2003interplay}.

In a conventional industrial setting, ambiguity in requirements is rarely problematic because it is actively discussed and resolved through goal-oriented conversations between stakeholders~\cite{ribeiro2020prevalence,de2010ambiguity,philippo2013requirement}. However, this level of refinement is difficult to achieve in OSS, where requesters and developers often lack direct communication channels. Although platforms like GitHub facilitate open discussions through comments, these conversations can become disorganized, unfocused, and cluttered with unrelated topics, making it challenging to reach a clear, actionable understanding of the request~\cite{heck2017framework}. 
% Furthermore, discussions may receive irrelevant input from other users, further obscuring the original intent of the feature request and hindering effective implementation~\cite{heck2017framework}. 

Prior research in requirements and software has tried to detect NL defects including ambiguity and incompleteness~\cite{fantechi2023rule, ezzini2022automated, zait2018addressing, ferrari2018detecting, mu2020nero, arora2019empirical, luitel2024improving, lian2024reqcompletion}. 
Ambiguity detection studies lack comprehensive categorization, focusing on narrow classifications, such as lexical or pragmatic ambiguity~\cite{fantechi2023rule, zait2018addressing, femmer2017rapid, ferrari2018detecting, yang2010automatic, gleich2010ambiguity, mu2020nero, seki2019detecting}. Incompleteness studies often rely on probabilistic models to predict the likely terms missing from sentences based on statistical patterns in the data~\cite{luitel2024improving}. However, by doing so, they can inadvertently misinterpret the requester's original intent. 
In addition, existing research focuses primarily on the detection of NL defects rather than their clarification and refinement~\cite{fantechi2023rule, ezzini2022automated, zait2018addressing, ferrari2018detecting, yang2010automatic, gleich2010ambiguity, mu2020nero, arora2019empirical, luitel2024improving}. 
While some studies attempt to resolve defects, their scope is often limited to specific patterns and domain models, addressing only certain types of missing information ~\cite{veizaga2024automated,arora2019empirical,ezzini2022automated}.

\subsection{Analysis of Issues Tracker in Github}
GitHub is one of the most widely used platforms for hosting OSS projects, providing tools for version control, collaboration, and project management. Among its core features is the ``Issues Tracker'' system, which allows developers and users to communicate, report bugs, propose new features, and engage in ongoing discussions about the software\cite{bissyande2013got}.

Previous research has explored the use of GitHub Issues for issue classification, project coordination, and community engagement, with particular attention to distinguishing between bugs, questions, and feature requests~\cite{kallis2019ticket, yang2023users, izquierdo2015gila, fiechter2021visualizing, hata2022github, bissyande2013got, li2022follow}. Kallis et al. introduce Ticket Tagger, a machine learning tool that categorizes issues (bug, feature, etc.) from titles/descriptions~\cite{kallis2019ticket}. Bissyande et al. investigate the adoption of issue trackers in terms of the projects that utilize them, the people that report issues, and the kind of issues that are usually reported~\cite{bissyande2013got}. Yang et al. present an empirical study that investigates the issues in open-source AI repositories to help developers understand problems during the process of using AI systems\cite{yang2023users}. Izquierdo et al. introduce a tool, which
generates a set of visualizations to facilitate the analysis of issues in a project depending on their label-based categorization~\cite{izquierdo2015gila}. Fiechter et al. present an interactive visual analytics tool to depict and analyze the relevant information pertaining to issue tales - visual narrative of the events and actors revolving around any GitHub issue~\cite{fiechter2021visualizing}.  

While these studies have improved our understanding of issue types and their management, the usage of conversational information or the comments in the issues involved in clarifying requests with NL defects remains largely underexplored.

\section{Approach}
\label{section:Approach}
\begin{figure*}
	\centering 
        \includegraphics[width=1\textwidth]{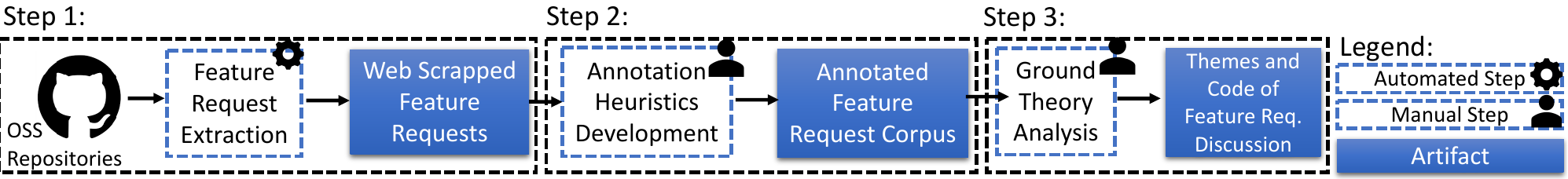}
	\centering
        \caption{Approach Overview}
	\label{fig:approach-overview1}
\end{figure*}

% We aim to address the following research questions (RQs) through a grounded-theory study of GitHub feature request issues.
% We address our research questions (RQs) through a grounded theory study of the feature request and discussion, employing systematic data collection and coding practices. Our goal is to uncover how developers engage and resolve clarification processes during feature request discussions in open source software (OSS) projects. 
We address our research questions (RQs) through a grounded theory study of the feature request discussions, employing systematic data collection and coding practices. Our approach uses primarily deductive coding based on our theoretical framework, while remaining open to inductive insights emerging from the data. 
%Our goal is to uncover how developers engage in and resolve clarification processes during feature request discussions in open source software (OSS) projects.
Figure~\ref{fig:approach-overview1} illustrates three main steps in our approach, which we discuss below.

% \textbf{RQ1:} How do developers engage in clarifying user-submitted feature requests in open source software (OSS) issue trackers?

\textbf{RQ1:} To what extent are feature requests prone to NL defects (Ambiguity and Incompleteness)?

% \textbf{RQ2:} How do developers engage in clarifying user-submitted feature requests containing NL defects in open source software (OSS) issue trackers?

% \textbf{RQ2:} What types of clarification do developers typically seek during feature request discussions (e.g., purpose, feasibility, implementation details, usage context)?
\textbf{RQ2:} How do developers seek to clarify NL defects in feature requests?

\textbf{RQ3:} Under what circumstances do developers opt not to seek clarification and instead choose to close or ignore feature requests?

\textbf{RQ4:} What practices support more effective clarification and lead to productive resolution of feature requests in OSS projects?

\subsection{Step 1: Feature Request Collection}
\label{ApproachStep1}
To construct a dataset of feature requests for our study, we select Mastodon and Signal, two well-known OSS projects on GitHub that serve as open-source alternatives to commercial platforms. We begin by scraping their GitHub repositories to identify issues labeled as ``\textit{feature}'' or ``\textit{feature request}'' collecting the corresponding issue IDs. Using the GitHub API, we retrieve the complete content for each issue, including the title, body, labels, status (open or closed), and associated comments. 
This process yields a dataset of 476 feature requests from both repositories together - 414 from Signal and 62 from Mastodon. 
%From this dataset, we randomly select 50 feature requests that have at least one comment to form our final analysis corpus. 
From this dataset, we randomly select 50 feature requests for our second step which is Feature Request Annotation (\ref{ApproachStep2}). 
% We analyze a total of 393 comments from the 50 feature requests selected. %\mitra{Add link to the data}
% The dataset is available online~\cite{AnonymizedRepo2025CrowdRE}.

%\footnote{https://anonymous.4open.science/r/Artifacts-CrowdRE2025-Submission/README.md}

\subsection{Step 2: Feature Request Annotation}
\label{ApproachStep2}
\begin{figure}
	\centering 
    \fbox{
	\includegraphics[width=0.45\textwidth]{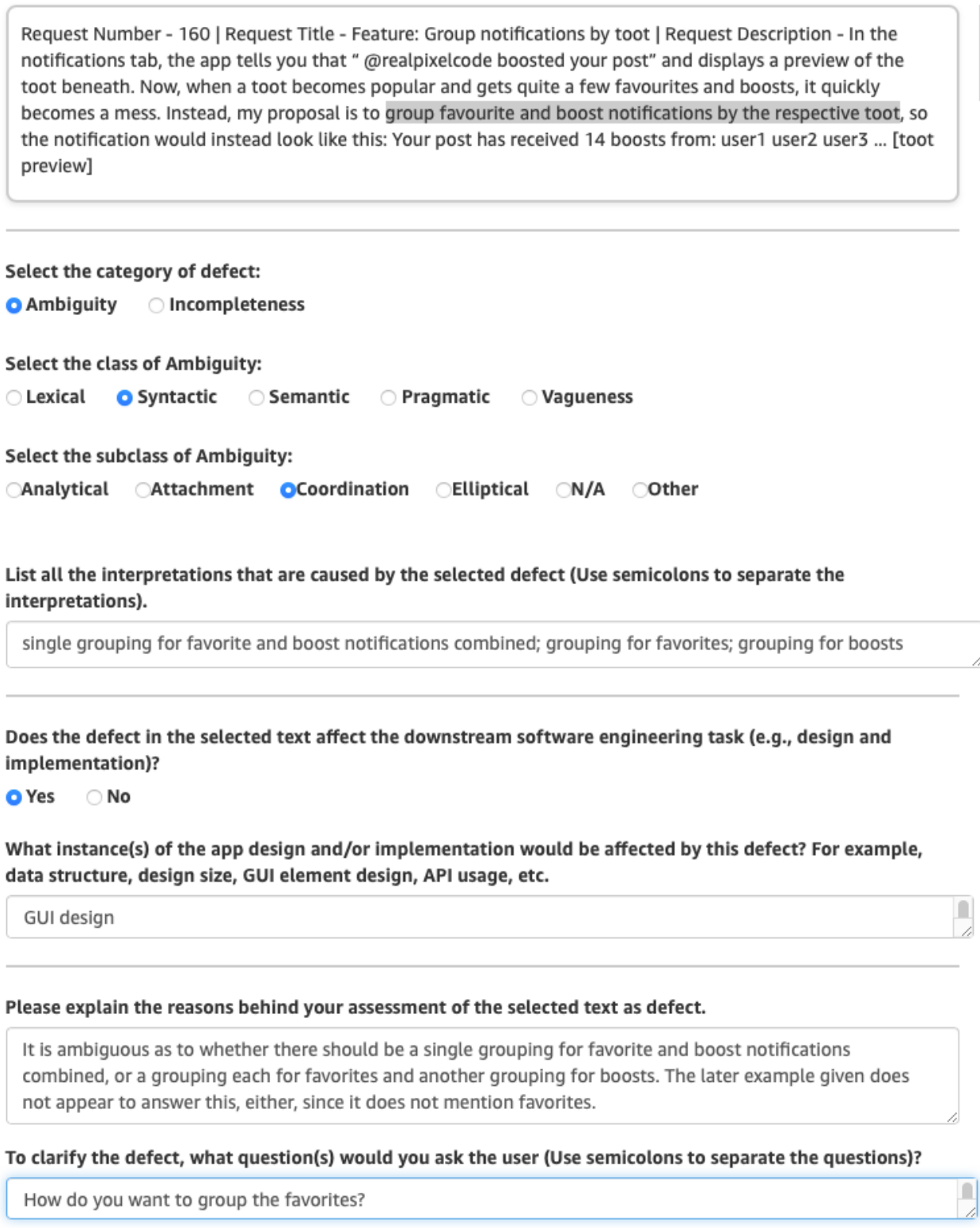}}
	\centering
        \caption{Annotation Tool}
	\label{fig:annotation}
    \vspace{-1em}
\end{figure}

To analyze the presence of NL defects in the 50 selected feature requests from Step 1, we create an annotation task where annotators are tasked to label NL defects (ambiguity and incompleteness), sub-classes (lexical, syntactic, semantic, pragmatic, and vagueness), interpretations, reasoning, and clarification questions (CQs). This annotated corpus will serve as our ground truth (GT) for the study. To ensure consistency with prior work on NL defect, we adopt ambiguity sub-classes from Berry et al~\cite{handbook2003contract}. Figure~\ref{fig:annotation} illustrates the tool designed for the annotation task. Initially, the authors familiarize themselves with handbook~\cite{handbook2003contract} definitions, independently analyze 15 feature requests, and meet to discuss findings. This process results in a heuristics document outlining the annotation procedures (see~\cite{AnonymizedRepo2025CrowdRE}). 

\begin{figure}
	\centering 
        \fbox{
	\includegraphics[width=0.45\textwidth]{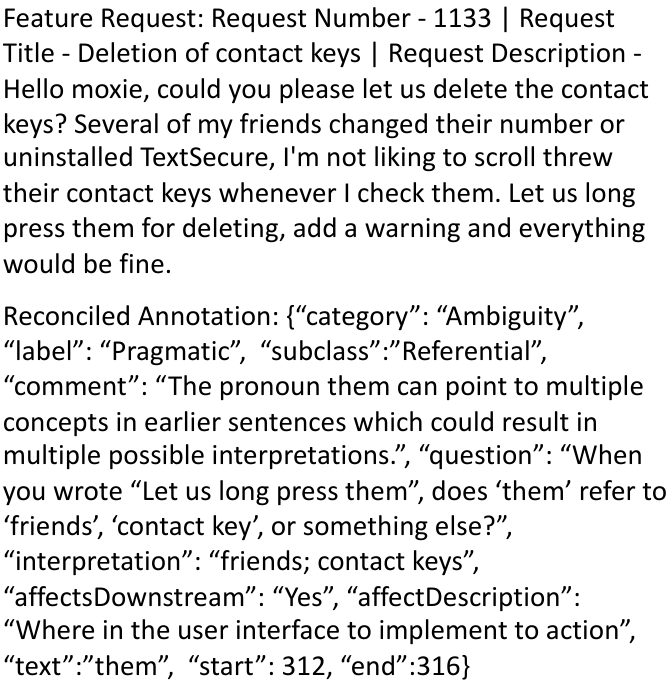}}
	\centering
        \caption{Annotation Example}
	\label{fig:annotationExample}
    \vspace{-1em}
\end{figure} 

Two annotators perform the annotations: Annotator A, a fluent non-native English speaker with an extensive background in SE, and Annotator B, a native English speaker with a background in linguistics and cybersecurity. Annotators performed analysis concurrently but independently of one another, meeting weekly to discuss whether any issues were encountered that required clarifying or modifying the heuristics document.

During annotation, if an ambiguity is detected, annotators highlight the relevant text segment and then specify sub-class of the ambiguity and offer their interpretations—since ambiguity requires at least two plausible interpretations~\cite{handbook2003contract}. For incompleteness, the annotators do not highlight text but instead describe what is missing from the feature request. Next, they assess whether the defect (i.e., ambiguity or incompleteness) could impact any downstream SE task, explaining how it might do so, and provide reasoning for their assessment. 
The annotators then provide CQs to help clarify the defect and refine the feature request. 

On average, both annotators spend 15 to 20 minutes per feature request. The Cohen Kappa~\cite{cohen1960coefficient} for Ambiguity is 0.61 and Incompleteness is 0.64 which indicate moderate agreement. Among the 50 requests, 16 contain no defects. A summary of the reconciled corpus is provided in Table~\ref{tab:statsForDefectsInGT} and ~\ref{tab:statsForDefectsInGT-Ambiguity}. An example of annotation of a reconciled feature is provided in Figure ~\ref{fig:annotationExample}.

\begin{table}[]
\caption{NL Defect Numbers}
\label{tab:statsForDefectsInGT}
\begin{tabular}{|l|l|l|l|}
\hline
 & \textbf{Ambiguity} & \textbf{Incompleteness} & \textbf{Both} \\ \hline
\textbf{\# of Feature Request} & 12 & 12 & 10 \\ \hline
\end{tabular}
\end{table}

\begin{table}[]
\caption{Ambiguity Sub Class Numbers}
\label{tab:statsForDefectsInGT-Ambiguity}
\small
\begin{tabular}{|l|c|c|c|c|c|}
\hline
 \textbf{Sub Classes} & \textbf{Lex.} & \textbf{Syn.} & \textbf{Sem.} & \textbf{Prag.} & \textbf{Vague.} \\ \hline
\textbf{\# of Feature Request} & 9 & 8 & 5 & 6 & 5 \\ \hline
\end{tabular}
\end{table}

%\mitra[inline]{If you have space, you can create a listing or figure to show the annotation for one of the requests in your dataset}
% \begin{lstlisting}[
%     label={lst:annotationDataExample},
%     caption=Annotation Example
%     ]
% Feature Request:
% Request Number - 1133 | Request Title - Deletion of contact keys | Request Description - Hello moxie, could you please let us delete the contact keys? Several of my friends changed their number or uninstalled TextSecure, I'm not liking to scroll threw their contact keys whenever I check them. Let us long press them for deleting, add a warning and everything would be fine.
% Reconciled Annotation:[{"category":"Ambiguity","label":"Pragmatic","subclass":"Referential","comment":"The pronoun \"them\" can point to multiple concepts in earlier sentences which could result in multiple possible interpretations.","question":"When you wrote \"Let us long press them\", does 'them' refer to 'friends', 'contact keys', or something else?","interpretation":"friends; contact keys","affectsDownstream":"Yes","affectDescription":"Where in the user interface to implement to action","text":"them","start":312,"end":316}]
% \end{lstlisting}

% corbin1990grounded
\subsection{Step 3: Analysis of Comments through Grounded Theory}\label{ApproachStep3}
To understand how developers clarify NL defects from discussion within issue requests, we adopt a structured manual coding approach on the corpus constructed from ~\ref{ApproachStep2}. For each of the 50 GT requests reviewed, we record the descriptive information, including the request number, request title, and request description, to maintain contextual understanding. We systematically examine the number of comments and the full text of the discussions to identify whether clarification questions are posed, whether the discussion is clear, the specific questions mentioned, and whether these questions contribute to clarifying or resolving the defect. We also evaluate whether the overall discussion helps to clarify the issue and identify the reason for the closure of the issue.

Our analysis primarily relies on deductive coding, where we apply categories derived from our research questions to systematically organize and compare data across requests (\cite{AnonymizedRepo2025CrowdRE}). These structured categories include ``Does the discussion have questions?', ``What are the questions mentioned if any?", ``Does the discussion help to clarify the request?" and ``Reason for closure?", allowing consistent data extraction and facilitating quantitative summarization of patterns observed in the dataset. This deductive approach ensures that our coding remains focused on the core aspects relevant to the study objectives i.e., understanding the patterns in the clarification process in the discussion.

In addition to this structured deductive coding, we also use inductive open coding while reviewing the requests, enabling the identification and recording of any emergent themes, unexpected clarification practices, or nuanced behaviors that are not captured by the categories mentioned earlier. The inductive coding is also used for finding categories for the reasons for feature request closure. This inductive coding complements the structured analysis by ensuring that novel insights and context-specific observations are incorporated into the findings, providing a richer and more comprehensive understanding of clarification dynamics in issue discussions. By combining these approaches, we achieve a rigorous analysis that systematically addresses our research questions while remaining open to new and meaningful patterns emerging from the data.

Each issue took approximately 18 minutes to analyze, allowing for careful, consistent application of grounded theory techniques across the dataset. This methodological design supports our aim to study how clarification is seen in practice and what contributes to its effectiveness in open-source development.

\section{Results and Discussion}
This section presents our findings from the grounded analysis and answers to the research questions discussed in~\ref{section:Approach}. We organize the discussion around our research questions, highlighting extent of NL defects on feature request (RQ1), what types of clarification they pursue (RQ2), when they refrain from doing so (RQ3), and what practices lead to clearer resolutions (RQ4).

\subsection{RQ1: NL Defects in Feature Requests}

Of the 50 feature requests analyzed, 34 contained NL defects, including ambiguity, incompleteness, or both. Among these, lexical and syntactic ambiguities were the most frequently observed types. Tables~\ref{tab:statsForDefectsInGT} and~\ref{tab:statsForDefectsInGT-Ambiguity} present detailed statistics on the identified defects.

The finding that over two-thirds of feature requests exhibit NL defects underscores a critical concern for requirements engineering practice. Such defects can result in misinterpretation of requirements, incorrect or inconsistent implementations, and increased communication overhead among stakeholders. This observation motivates our second research question, which investigates how developers seek clarification on these feature requests with NL defects.

% \subsection{RQ2: User-Developer Engagement in Clarifying Feature Requests}
% Our analysis finds that developers generally adopt a practical, goal-oriented position when responding to user-submitted feature requests. Instead of focusing on linguistic details or grammatical quality, their clarifying efforts often aim to determine whether a feature aligns with the project’s goals or user needs. Clarification is rarely about textual interpretation—it is about understanding intent and feasibility.

\subsection{RQ2: Types of Clarification Developers Seek}
\begin{figure}
	\centering 
        \fbox{
	\includegraphics[width=0.45\textwidth]{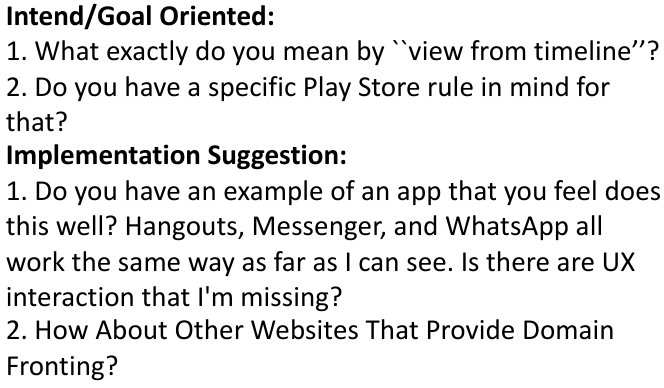}}
	\centering
        \caption{Clarifying Questions asked by Developers}
	\label{fig:clarifyingExample}
    \vspace{-1em}
\end{figure}

% We observe that clarifying questions, when asked by developers, typically revolve around the user’s intent, motivations, or desired outcomes. 
Developers seek to understand what users are trying to achieve, rather than exploring into implementation specifics unless the user initiates that level of detail or the textual details which relates to NL defects. Examples for such are shown in Figure ~\ref{fig:clarifyingExample}. 

\begin{figure}
	\centering 
        \fbox{
	\includegraphics[width=0.45\textwidth]{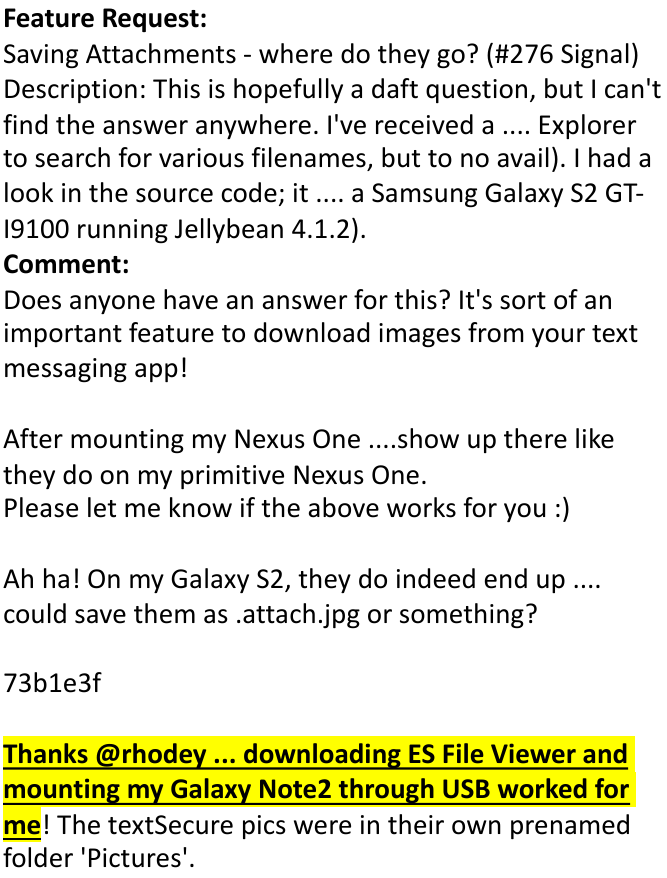}}
	\centering
        \caption{Clear Discussion for Clarification}
	\label{fig:ClearConversationExample}
    \vspace{-1em}
\end{figure}

% Developers rarely asked clarifying questions in regards to NL defects. In 39 out of the 50 analyzed issues, developers did not ask any clarifying questions despite our GT having clarification questions for the request.  
% Lack of discussion around the textual quality of the request was observed even though the requests contain NL defects such as ambiguity and incompleteness. 
Developers rarely asked clarifying questions regarding NL defects. In 39 out of the 50 issues analyzed, developers did not seek any clarification, even though our ground truth identified clarification questions for these requests. This indicates a lack of discussion around the textual quality of the requests, despite the presence of NL defects such as ambiguity and incompleteness. Developers either implemented the request directly or closed it without further discussion. Instead of seeking clarification, developers often responded by directing users to existing features that addressed their needs, indicating a preference for reusing existing solutions rather than implementing new features. Additionally, in some cases, other users – not the developers – provided the necessary clarifications or suggested workarounds for the requested features, as shown in Figure ~\ref{fig:ClearConversationExample}.
%Also notable is the lack of discussion around the textual quality of the request even though the requests contain NL defects such as ambiguity and incompleteness. Developers very rarely commented on ambiguity or incompleteness, suggesting an implicit expectation that users submit clear, actionable requests.

There are also only very few to none instances in the analyzed discussions that developers or users explicitly address how the proposed feature should be integrated into the graphical user interface (GUI). Conversations lack details about where the feature would appear, how users would interact with it, or how it would align with existing visual designs and workflows. This absence of GUI-level consideration may hinder implementation clarity—particularly for features that rely heavily on user interaction—and can result in mismatches between user expectations and the final design. 
Moreover, developers rarely ask about technical implementation unless specifically mentioned by the user or accompanied by a code contribution. This implies that clarification is generally strategic (does the feature fit?), rather than technical (how would it work?) during early discussion phases. 

% \begin{lstlisting}[
%     label={lst:clarifyingQues},
%     caption=Clarifying questions asked by developers
%     ]
% Intend/Goal Oriented:
%     1. What exactly do you mean by ``view from timeline''?
%     2. Do you have a specific Play Store rule in mind for that?
% Implementation Suggestion:
%     1. Do you have an example of an app that you feel does this well? Hangouts, Messenger, and WhatsApp all work the same way as far as I can see. Is there are UX interaction that I'm missing?
%     2. How About Other Websites That Provide Domain Fronting?
% \end{lstlisting}

\subsection{RQ3: Conditions for Non-Clarification Closure}

There are situations where developers choose not to engage in clarification and instead move directly to closing the request immediately without implementation. Table ~\ref{tab:closureReason} presents the reasons for closure and frequency. Most of the time, developers state why they will not proceed with a request and close the thread without further clarification. However, there are instances where developers close the thread without providing any reason. 
%Developers for most of the times state why they will not proceed with the request and close the thread which leads to no clarification but there are instances where no reasons were provided by the developers for the closure. 

\begin{table*}[]
\centering
\caption{Reason for closure}
\label{tab:closureReason}
\begin{tabular}{|l|l|l|l|l|l|}
\hline
 & \textbf{Existing Feature} & \textbf{Duplicate Request} & \textbf{Privacy Concern} & \textbf{No Reason} & \textbf{Clean up/Bug Report} \\ \hline
\textbf{\# of Feature Req.} & 4 & 7 & 3 & 5 & 9 \\ \hline
\end{tabular}
\end{table*}

\begin{figure}
	\centering 
        \fbox{
	\includegraphics[width=0.45\textwidth]{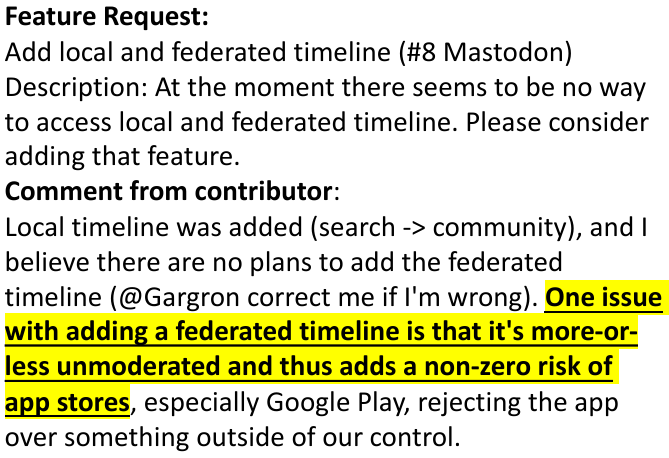}}
	\centering
        \caption{Feature Request Closed for Security Concern}
	\label{fig:SecurityConcern}
    \vspace{-2em}
\end{figure}

The first reason for closure is the presence of a security vulnerability or privacy issue as shown in Figure \ref{fig:SecurityConcern}, either inherent in the request itself or in its potential consequences. Developers tend to treat such issues with high sensitivity and close them, often without seeking additional input from the user. Three out of 50 requests from the dataset were closed with Privacy or Legal concern. The developers did provided how the request would cause a legal concern or bring security vulnerability to the software before closing.  

% \begin{lstlisting}[
%     label={lst:privacyConcern},
%     caption=Feature Request closed for security concern
%     ]
% Feature Request:
%     Add local and federated timeline (#8 Mastodon)
%     Description: At the moment there seems to be no way to access local and federated timeline. Please consider adding that feature.
% Comment:
%     Local timeline was added (search -> community), and I believe there are no plans to add the federated timeline (@Gargron correct me if I'm wrong). One issue with adding a federated timeline is that it's more-or-less unmoderated and thus adds a non-zero risk of app stores, especially Google Play, rejecting the app over something outside of our control. 
% \end{lstlisting}

\begin{figure}
	\centering 
        \fbox{
	\includegraphics[width=0.45\textwidth]{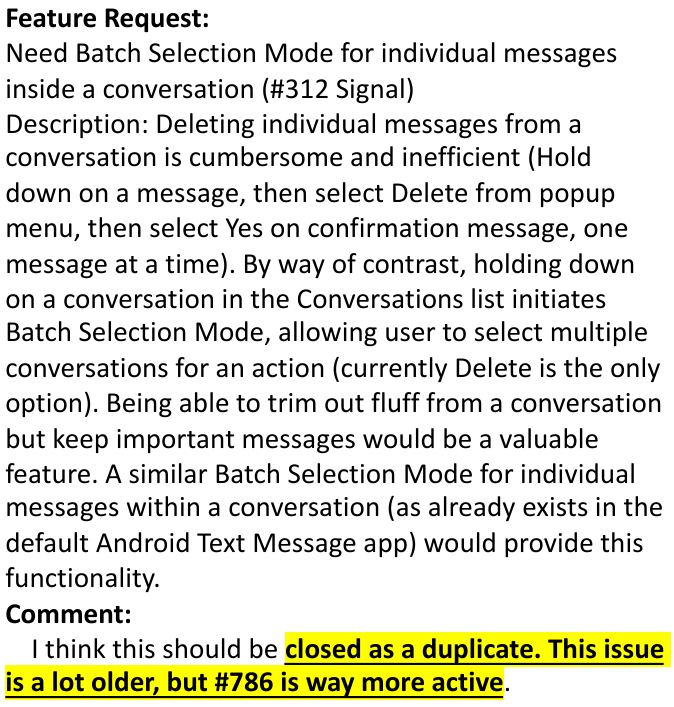}}
	\centering
        \caption{Feature Request closed marked as Duplicate Request}
	\label{fig:dulicateRequest}
    \vspace{-1em}
\end{figure}

Duplicate request is another reason to skip the requester's clarification and lead to immediate closure. When developers identify that a similar or identical feature has already been submitted or resolved (positively or negatively), they typically reference the earlier thread and close the new issue, as shown in Figure ~\ref{fig:dulicateRequest}. This indicates a preference for issue consolidation, reflecting the community’s emphasis on efficiency and reducing redundancy. Seven instances out of 50 requests were closed as duplicate request.
% \begin{lstlisting}[
%     label={lst:duplicateRequest},
%     caption=Feature Request closed marked as Duplicate Request
%     ]
% Feature Request:
%     Need Batch Selection Mode for individual messages inside a conversation (#312 Signal)
%     Description: Deleting individual messages from a conversation is cumbersome and inefficient (Hold down on a message, then select Delete from popup menu, then select Yes on confirmation message, one message at a time).

%     By way of contrast, holding down on a conversation in the Conversations list initiates Batch Selection Mode, allowing user to select multiple conversations for an action (currently Delete is the only option).

%     Being able to trim out fluff from a conversation but keep important messages would be a valuable feature. A similar Batch Selection Mode for individual messages within a conversation (as already exists in the default Android Text Message app) would provide this functionality.
% Comment:
%     I think this should be closed as a duplicate. This issue is a lot older, but #786 is way more active.

%     agree, thanks!
% \end{lstlisting}

Moreover, five instances of the 50 analyzed requests had no explicit reason provided for why the request thread was closed. 10 out of the 50 requests were also closed as bug reports/clean up, even though they were labeled as feature requests. These were closed because if app versions, updates, and bugs.

These closure decisions highlight that developers look not just at technical considerations, but also at project alignment, risk assessment, and workload management. In cases where clarification is perceived as unnecessary or unproductive, developers opt to preserve their time and focus by terminating the request early.

\subsection{RQ4: Practices That Enhance Clarification and Resolution}

The second research question explores what practice(s) or condition(s) facilitate more productive clarification processes and improve the overall outcome of feature request handling.

\begin{figure}
	\centering 
        \fbox{
	\includegraphics[width=0.45\textwidth]{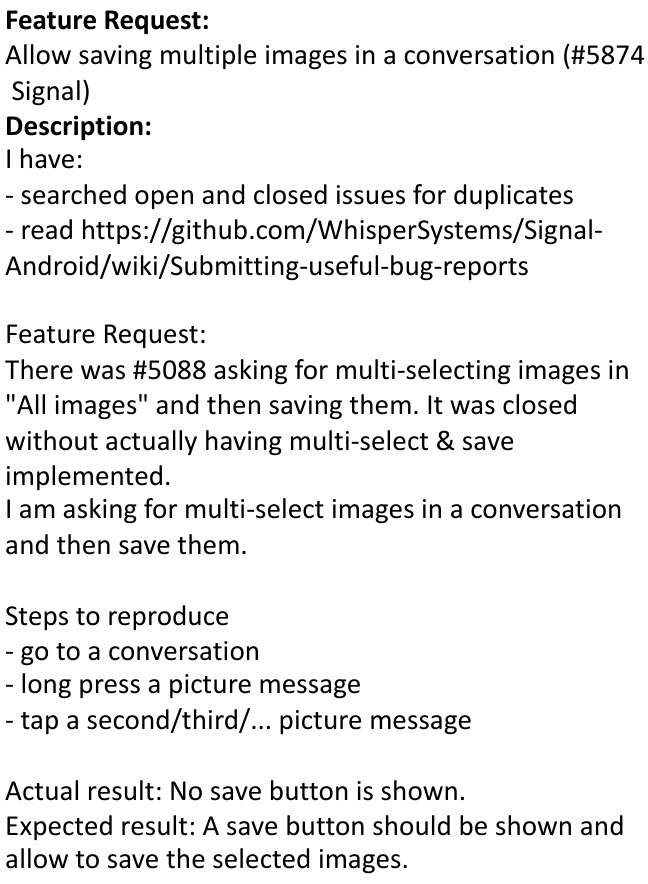}}
	\centering
        \caption{Feature Request with template}
	\label{fig:templateForRequest}
    \vspace{-2em}
\end{figure}

Our findings indicate that developers are significantly more responsive and collaborative when users submit requests that are well-structured and clearly articulated, as shown in Figure ~\ref{fig:templateForRequest}\footnote{Ellipses (...) are used to indicate omitted content for brevity.}. Specifically, requests that clearly state the problem, justify the need for the feature, and provide at least a rough idea of the desired behavior or implementation approach tend to elicit more feedback and interaction from developers. Of the 50 requests analyzed, the seven that followed a structured template defined their intent effectively and required minimal clarification. In contrast, the remaining requests, which were written in a free-form style, often suffered from inconsistency and missing details. Furthermore, users who included mock-ups, example code snippets, or links to similar implementations typically received faster and more positive responses from developers.

The presence of implementation cues and suggestion — even if informal or incomplete — appears to lower the cognitive load for developers and reduces ambiguity about the user’s expectations. It also demonstrates user commitment and technical awareness, which developers often reward with more attention and effort.

% 17 requests only implemented out of 50
One solution for maintaining a well structured and clear feature request in issues is the adoption of standardized issue templates for feature requests and making it required, as shown in Figure~\ref{fig:templateForRequest}. Incomplete adherence to the template or submission of free-form requests introduces more ambiguity and overhead for maintainers, who must spend additional time clarifying requirements and reproducing the described functionality~\cite{sulun2024empirical}. Figure~\ref{fig:MastodonIssueTracker} shows the issue tracker for Mastodon that allows users to write feature request without template. This might bring more inconsistency since the user might provide defective requests. By enforcing a consistent structure, templates improve communication efficiency, reduce misunderstandings, and ultimately streamline the feature development process. A sample of a standardized required template could include:
\begin{figure}
	\centering 
	\includegraphics[width=0.5\textwidth]{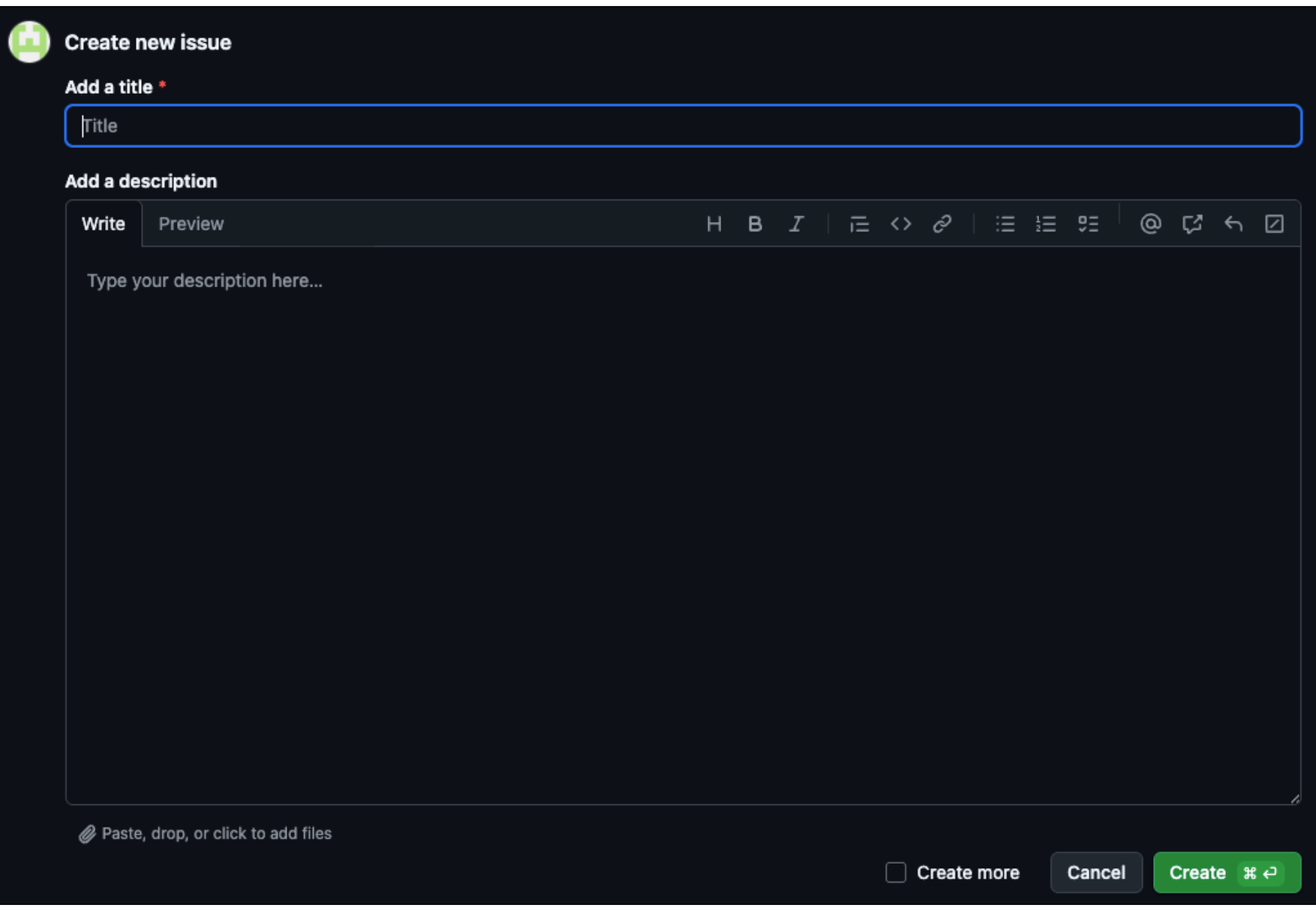}
	\centering
    \caption{Mastodon Issue Tracker}
	\label{fig:MastodonIssueTracker}
    \vspace{-2em}
\end{figure} 

\begin{itemize}
    % \item A clear description of the feature
    \item The motivation or problem being solved
    \item Expected behavior and use cases
    \item Optional: Code snippets or examples from other related software
\end{itemize}

Such templates would reduce NL defects, streamline clarification, and allow developers to more quickly assess the viability and priority of requests. Moreover, enforcing templates promote consistency across submissions, making it easier for maintainers to compare and resolve issues.

\subsection{Insights and Implications}
Our findings suggest several critical insights with direct implications for both OSS communities and Requirements Engineering communities:
(1) \textbf{Clarification Focus is Goal-Oriented:} Developers in OSS projects tend to prioritize understanding the overarching intent and feasibility of a feature request rather than resolving its linguistic ambiguities or incompleteness. This suggests that clarification efforts are driven by strategic project alignment rather than textual correctness. For RE researchers, this highlights the importance of designing tools and interventions that support intent clarification over low-level textual disambiguation. 
(2) \textbf{Limited Explicit Clarification:} Despite the prevalence of NL defects, explicit clarifying questions are rare. Developers often proceed to decision-making without resolving ambiguities or missing details. This has two implications: (i) users bear the burden of providing clear and complete requests; (ii) requests with defects risk being closed without implementation or misinterpreted, potentially lowering user satisfaction and leading to feature rejection.
(3) \textbf{Implications for Tooling:} Existing automated defect detection approaches focus mainly on identifying ambiguities and incompleteness without facilitating effective clarification. Our results suggest that tools should evolve to support the formulation of context-relevant clarifying questions, enabling developers to elicit missing information efficiently and reducing cognitive load during issue resolution.
(4) \textbf{Towards template reinforcement:} The presence of structured, template-based requests correlated with clearer understanding, reduced need for clarification, and more constructive developer responses. OSS maintainers should enforce the use of feature request templates that explicitly elicit user goals, motivation, expected behavior, and optional implementation suggestions. For RE practice, standardized templates can reduce NL defects at the requirements elicitation stage and enhance communication efficiency.

\section{Threats to validity}\label{sec:threats}
This study has several threats to validity that should be considered when interpreting the results.
(1) Construct Validity: Our annotation process for NL defects relied on definitions adapted from prior literature and subjective judgments by annotators. Although we employed two annotators with complementary expertise and used reconciliations to reduce bias, interpretations of ambiguity and incompleteness remain inherently subjective, potentially affecting consistency. (2) Internal Validity: The grounded theory analysis was conducted systematically; however, potential might bias exists in open coding and theme identification. While structured deductive categories ensured focus on research questions, the coding decisions might have overlooked subtle nuances in discussions that did not fit predefined categories. (3) External Validity: The dataset comprises feature requests from only two OSS projects: Signal and Mastodon, which limits generalizability to other types of software projects or domains. These projects may have distinct community cultures and moderation practices that influence clarification dynamics differently compared to system software, developer tools, or enterprise OSS projects.

\vspace{-1em}
\section{Conclusion \& Future Directions}\label{sec:conclusion}
In this paper, we investigated how developers seek clarification of user-submitted feature requests in open-source issue trackers, focusing on GitHub repositories for Mastodon and Signal. Our grounded analysis revealed that explicit clarification is relatively rare; developers often prioritize alignment with project goals over resolving textual ambiguity. When clarification does occur, it centers on understanding user intent and practical feasibility rather than technical specifics. We found that well-articulated requests, those that clearly state a problem, motivation, and expected behavior, are more likely to receive constructive responses. This highlights the importance of structured communication in facilitating productive collaboration. For future work, we aim to expand the dataset and review additional issues from various other open-source software projects. Further, we also look to conduct interviews and surveys to understand developers' perspectives and to check the feasibility of enforcing the template-based request.

\section{Acknowledgment}\label{sec:acknowledgment}
This work is supported by NSF award \#2318915.

\bibliographystyle{IEEEtran}
\bibliography{reference}

\end{document}